\documentclass[prl,twocolumn,aps,showpacs,superscriptaddress]{revtex4-1}
\usepackage{amsmath}
\usepackage{dcolumn}
\usepackage{graphicx}

\usepackage{bm}
\usepackage{color}
\usepackage{ulem}

\newcommand{\GTS}{GaTa$_4$Se$_8$}

\begin{document}

\title{Novel $J_{\rm{eff}}$=3/2 Metallic Phase and Unconventional Superconductivity in GaTa$_4$Se$_8$}

\author{Min Yong Jeong}
\affiliation{Department of Physics, Korea Advanced Institute of Science and Technology, Daejeon 34141, Korea }

\author{Seo~Hyoung Chang}
\affiliation{Department of Physics, Chung-Ang University, Seoul 06974, South Korea}

\author{Hyeong Jun Lee}
\affiliation{Center for Theoretical Physics of Complex Systems, Institute for Basic Science (IBS), Daejeon 34126, Korea }

\author{Jae-Hoon Sim} 
\affiliation{Department of Physics, Korea Advanced Institute of Science and Technology, Daejeon 34141, Korea }
\affiliation{CPHT, CNRS, Ecole Polytechnique, Institut Polytechnique de Paris, F-91128 Palaiseau, France}

\author{Kyeong Jun Lee}
\affiliation{Department of Physics, Chung-Ang University, Seoul 06974, South Korea}

\author{Etienne Janod} 
\affiliation{Institut des Mat\'eriaux Jean Rouxel (IMN), Universit\'e de Nantes, CNRS, 2 Rue de la Houssini\`ere, BP32229, 44322 Nantes cedex 3, France}

\author{Laurent Cario} 
\affiliation{Institut des Mat\'eriaux Jean Rouxel (IMN), Universit\'e de Nantes, CNRS, 2 Rue de la Houssini\`ere, BP32229, 44322 Nantes cedex 3, France}

\author{Ayman Said}
\affiliation{Advanced Photon Source, Argonne National Laboratory, Argonne, IL 60439, USA}

\author{Wenli Bi}
\affiliation{Department of Physics, University of Alabama at Birmingham, Birmingham, AL, 35294, USA}

\author{Philipp Werner} 
\affiliation{Department of Physics, University of Fribourg, 1700 Fribourg, Switzerland}

\author{Ara Go} \email{arago@jnu.ac.kr}
\affiliation{Center for Theoretical Physics of Complex Systems, Institute for Basic Science (IBS), Daejeon 34126, Korea }
\affiliation{Department of Physics, Chonnam National University, Gwangju 61186, Korea}

\author{Jungho Kim} \email{jhkim@anl.gov}
\affiliation{Advanced Photon Source, Argonne National Laboratory, Argonne, IL 60439, USA}

\author{Myung Joon Han} \email{mj.han@kaist.ac.kr}
\affiliation{Department of Physics, Korea Advanced Institute of Science and Technology, Daejeon 34141, Korea }

\begin{abstract}
By means of density functional theory plus dynamical mean-field theory (DFT+DMFT) calculations and resonant inelastic x-ray scattering (RIXS) experiments, we investigate the high-pressure phases of the spin-orbit-coupled $J_{\rm{eff}}=3/2$ insulator GaTa$_4$Se$_8$. Its metallic phase, derived from the Mott state by applying pressure, is found to carry $J_{\rm{eff}}=3/2$ moments. The characteristic excitation peak in the RIXS spectrum maintains its destructive quantum interference of $J_{\rm{eff}}$ at the Ta $L_2$-edge up to 10.4 GPa. Our exact diagonalization based DFT+DMFT calculations including spin-orbit coupling also reveal that the $J_{\rm{eff}}=3/2$ character can be clearly identified under high pressure. These results establish the intriguing nature of the correlated metallic magnetic phase, which represents the first confirmed example of $J_{\rm{eff}}$=3/2 moments residing in a metal. They also indicate that the pressure-induced superconductivity is likely unconventional and influenced by these $J_{\rm{eff}}=3/2$ moments. Based on a self-energy analysis, we furthermore propose the possibility of doping-induced superconductivity related to a spin-freezing crossover. 
\end{abstract}

\maketitle

{\it Introduction --}
Identifying and characterizing the phases and phase transitions of materials is a central theme of condensed matter physics. The discovery of a new type of phase often requires theoretical analyses of its essential nature, as well as clarifications of the relationship to other known phases and the possible transitions into nearby phases. As a well-known example, unconventional metal states in cuprate phase diagram hold many mysteries \cite{fradkin_colloquium_2015,norman_pseudogap_2005,lee_doping_2006,keimer_physics_2017}. Being clearly different from a Fermi liquid, these anomalous metallic phases can be a precursor or competitor of unconventional superconductivity \cite{basov_electrodynamics_2011,scalapino_common_2012,kapitulnik_colloquium_2019}.

The lacunar spinel GaM$_4$X$_8$ (M=V, Nb, Ta, Mo; X=S, Se, Te) is a fascinating class of materials which exhibits multiferroic, skyrmion, and resistive switching phenomena \cite{kezsmarki_ne-type_2015,ruff_multiferroicity_2015,fujima_thermodynamically_2017,ruff_polar_2017,geirhos_orbital-order_2018,guiot_control_2011,dubost_resistive_2013}. GaTa$_4$Se$_8$, in particular, has been highlighted as an interesting example that undergoes a paramagnetic Mott insulator to metal transition (IMT) under pressure \cite{abd-elmeguid_transition_2004,pocha_crystal_2005,ta_phuoc_optical_2013,camjayi_first-order_2014}. Furthermore, recent studies elucidated the significant effect of spin-orbit coupling (SOC) and showed that its ground state carries spin-orbit entangled (so-called) $J_{\rm eff}$=3/2 moments \cite{kim_novel_2008,kim_spin-orbital_2014,jeong_direct_2017}, which is the first confirmed example of this kind.
Considering the observed IMT followed by a superconducting transition as a function of pressure, the identification of the $J_{\rm eff}$=3/2 Mott phase at ambient conditions immediately generates a series of important questions: If the metallic phase is a conventional Fermi liquid, it is a more or less trivial case, and the superconductivity observed at higher pressures is also likely of the conventional type. On the other hand, if it is a correlated metal which still hosts $J_{\rm eff}=3/2$ moments, it can be regarded as a new type of metallic phase, and the observed superconductivity is more likely to be unconventional.\\
\begin{figure}[ht]
	\includegraphics[width=0.45\textwidth]{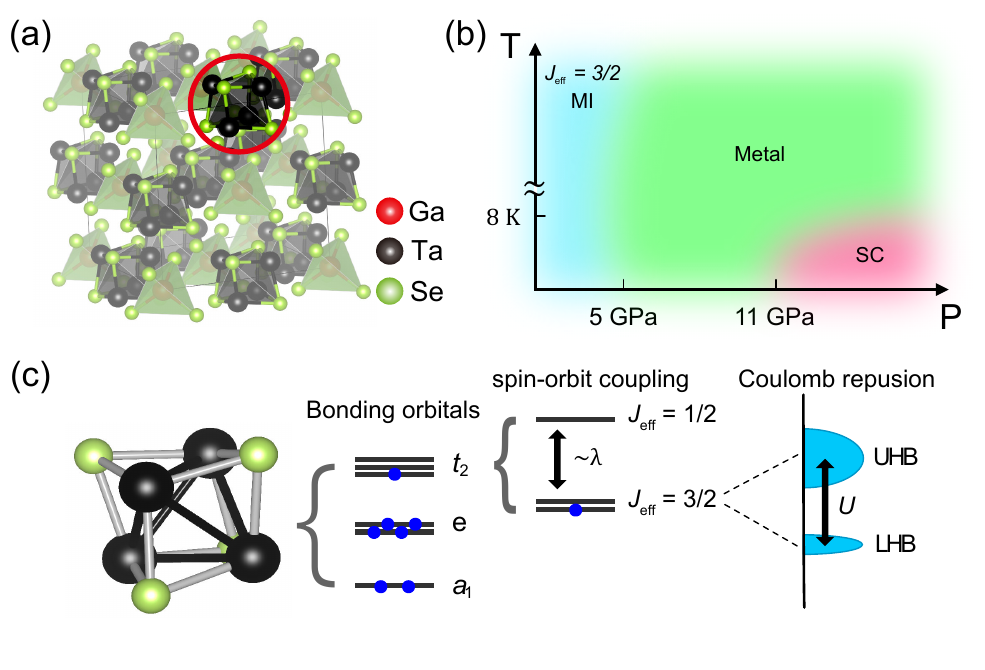}
	\caption{(a) Crystal structure of GaTa$_4$Se$_8$. Red, black and green spheres represent the Ga, Ta and Se atoms, respectively. (b) A schematic pressure-temperature phase diagram of GaTa$_4$Se$_8$. The cyan, green and red colored region represents the $J_{\rm{eff}}=3/2$ Mott insulating, metallic and superconducting phases, respectively. It should be noted that the phase boundary lines have not yet been well identified due to the lack of experimental information. (c) Schematic electronic structure near $E_F$ which is dominated by molecular orbital states of the higher-lying $t_2$ and lower-lying $e$ and $a_1$ type. The $t_{2}$ levels are further split into $J_{\rm{eff}}=1/2$ and $3/2$ by SOC. At ambient pressure, the Mott gap is stabilized by the on-site Coulomb interaction $U$.
	}
	\label{crystal}
\end{figure}
In this Letter, we try to elucidate the nature of the pressure-induced metallic phase which emerges out of the Mott insulator without doping. By means of resonant inelastic x-ray scattering (RIXS) experiments and density functional theory plus dynamical mean-field theory (DFT+DMFT) calculations, we investigate its detailed electronic and magnetic properties. We find that the characteristic $L_3$ peak is clearly observed even in the metallic regime while the forbidden $L_2$ peak is absent. This observation together with the simulation results clearly identifies the novel metallic state with $J_{\rm eff}=3/2$ magnetic moments. We discuss its implications regarding the superconductivity at higher pressure. Finally, we explore another intriguing possibility in this material. Our self-energy analysis shows that electron doping can induce a spin-freezing crossover, a phenomenon which has been previously linked to unconventional superconductivity \cite{hoshino_superconductivity_2015}. 
These results will hopefully stimulate experimental efforts to clarify the properties of this material under chemical or other types of doping.


{\it Electronic structure and insulator-metal-superconductor transition --}
GaTa$_4$Se$_8$ is composed of well-separated GaSe$_4$ and Ta$_4$Se$_4$ molecular clusters as shown in Fig.~\ref{crystal}(a). The Fermi level ($E_F$) is dominated by  $t_2$ molecular orbitals which are derived from Ta-$t_{2g}$ atomic orbitals \cite{pocha_electronic_2000,abd-elmeguid_transition_2004,pocha_crystal_2005,camjayi_localised_2012,ta_phuoc_optical_2013,dubost_resistive_2013,camjayi_first-order_2014,kim_spin-orbital_2014,jeong_direct_2017}. On top of the spin-orbit splitted molecular $J_{\rm{eff}}=3/2$ quartet and $J_{\rm eff}=1/2$ doublet, the on-site Coulomb interaction ($U$) induces a Mott gap in the quater-filled $J_{\rm{eff}}=3/2$ bands; see Fig.~\ref{crystal}(c) \cite{kim_spin-orbital_2014,jeong_direct_2017}. This spin-orbit entangled molecular $J_{\rm{eff}}=3/2$ Mott phase was first predicted by DFT+SOC$+U$ calculations \cite{kim_spin-orbital_2014} and then confirmed by RIXS experiments \cite{jeong_direct_2017}. Here the `on-site' Coulomb repulsion $U$ represents the interaction within molecular $t_2$ orbitals rather than atomic Ta orbitals \cite{pocha_electronic_2000,pocha_crystal_2005,camjayi_first-order_2014}.

Largely unexplored are the IMT and the metal-to-superconductor transition, both of which are induced by applying pressure (without doping; see Fig.~\ref{crystal}(b)) \cite{abd-elmeguid_transition_2004,pocha_crystal_2005,ta_phuoc_optical_2013,camjayi_first-order_2014}. The pressure-dependent crystal structure data \cite{pocha_crystal_2005} indicate that the Mott IMT is caused by the increased hopping integrals between the Ta$_4$Se$_4$ molecular units. This bandwidth controlled IMT was studied based on the three-orbital Hubbard model within DMFT-QMC (quantum Monte Carlo) \cite{camjayi_first-order_2014}. However, the effect of SOC has not been taken into account and therefore the $J_{\rm{eff}}=3/2$ state could not be realized.

\begin{figure}[t]
	\includegraphics[width=0.45\textwidth]{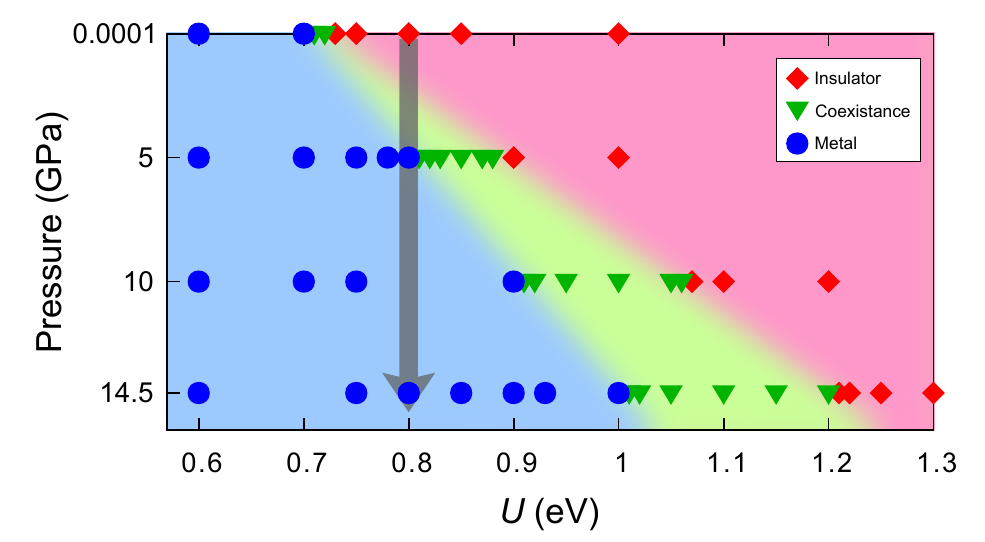}
	\caption{Calculated phase diagram for \GTS~as a function of $U$ and pressure within DFT+DMFT+SOC (at zero temperature). Blue circles, red diamonds and green triangles represent the calculated points corresponding to the metallic, insulating, and coexistence phase, respectively. The realistic value of $U\approx0.8$ eV is depicted by a gray arrow.
	}
	\label{phase_diagram}
\end{figure}

{\it DFT+DMFT phase diagram: The effect of SOC --}
With this motivation, we first performed DFT+DMFT calculations with SOC 
(See Supplemental Materials for computation details \cite{supplementary_materials}). The calculated phase diagram is presented in Fig.~\ref{phase_diagram}. The red and blue colored regions represent the insulating and metallic phases, respectively. We note that at $U > 0.7$ eV, pressure can always induce the transition and that the critical value of $U_c$ is gradually increased as pressure increases.

By including SOC, the calculated phase diagram shows a good quantitative agreement with the experiments. Considering the neglected frequency dependence of $U$ in the DMFT procedure, we expect that the realistic effective interaction strength is slightly larger than the constrained random phase approximation (cRPA) value of $U_{\rm cRPA}=0.7$ eV; $U\approx 0.7$-$0.9$ eV \cite{casula_low-energy_2012}. With $U=0.8$ eV, the IMT occurs at $P\approx 5$ GPa as shown in Fig.~\ref{phase_diagram}. This is in good agreement with previous experimental data reporting a critical pressure $P_c$ of $5$-$7$ GPa \cite{ta_phuoc_optical_2013,camjayi_first-order_2014}.
It is important to note that, in the previous DMFT calculations (without SOC), the critical $U_c$ of 1.2 eV \cite{camjayi_first-order_2014} is significantly larger than our value. Also, if we follow  Ref.~\onlinecite{camjayi_first-order_2014} and identify the calculated coexistence region with the hysteresis region observed at intermediate pressure in the resistivity measurement \cite{camjayi_first-order_2014}, our results are in even better agreement with the experiment.
Hence, without the effect of SOC, the experimental phase boundary cannot be well reproduced and the  $J_{\rm{eff}}=3/2$ moments are not formed.

{\it Metallic J$_{\rm eff}$=3/2 states: RIXS experiment --}
The direct evidence of the novel $J_{\rm{eff}}=3/2$ Mott phase at ambient pressure came from RIXS  \cite{jeong_direct_2017}. As an element-specific photon-in and photon-out measurement using dipole transitions between Ta 5$d$ and 2$p_{3/2}$ ($L_{3}$) or 2$p_{1/2}$ ($L_{2}$), RIXS was able to detect and compare the excitation spectra at both edges. The compelling evidence for $J_{\rm{eff}}=3/2$ was the presence and the absence of a $\sim$1.3 eV peak at $L_{3}$ and $L_{2}$, respectively, which is directly based on the quantum mechanical selection rules \cite{jeong_direct_2017}. Here we adopt the same approach to probe $J_{\rm{eff}}=3/2$ moments in the metallic regime and perform the pressure-dependent RIXS measurements (See Supplemental Materials for experimental details \cite{supplementary_materials}).

\begin{figure}[t]
	\includegraphics[width=0.45\textwidth]{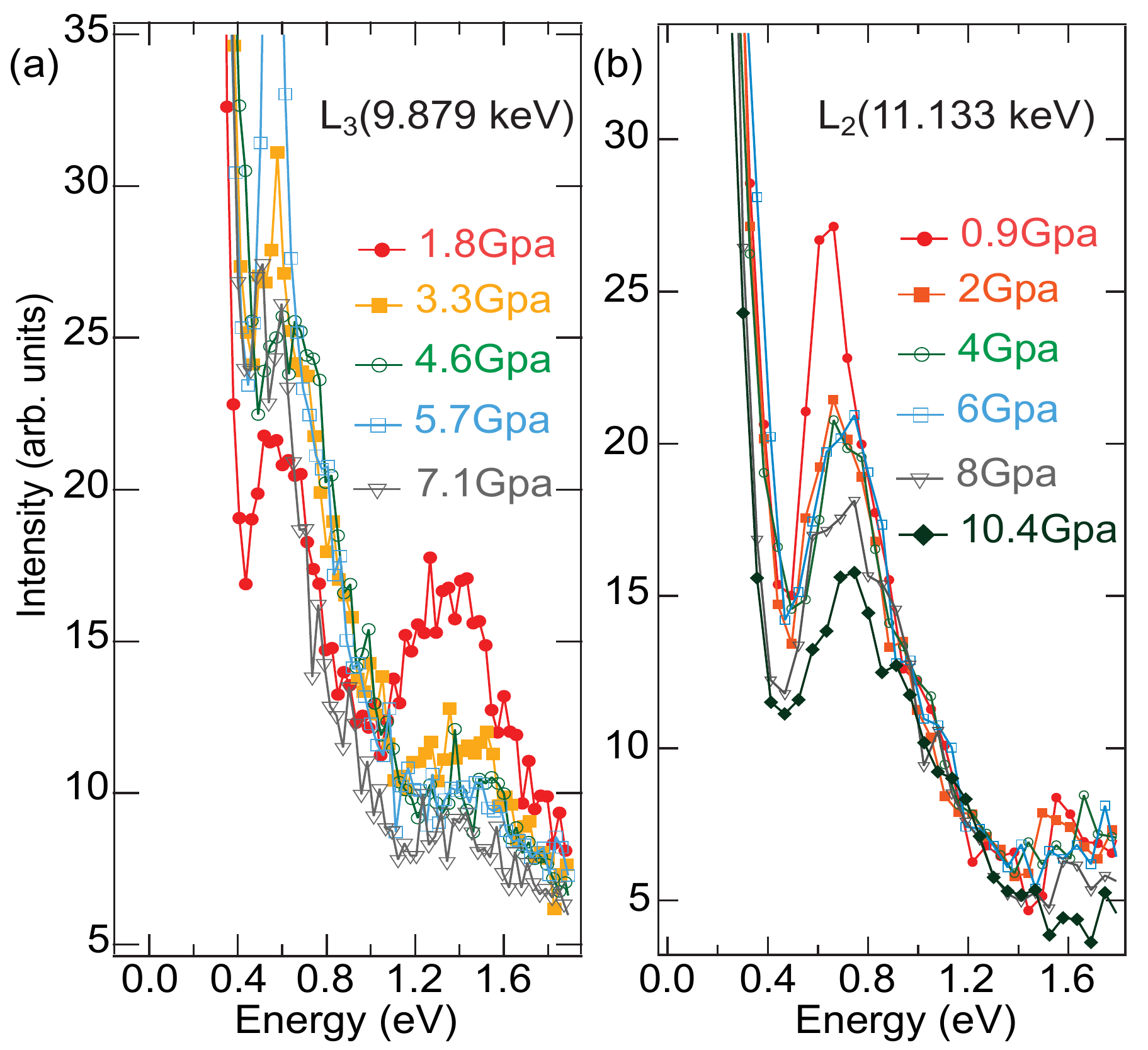}
	\caption{(a, b) Pressure-dependent RIXS data at the (a) $L_3$ and (b) $L_2$ edge. The different symbols and colors represent different pressure values.
	}
	\label{RIXS}
\end{figure}

Figure~\ref{RIXS}(a) shows the RIXS spectra at the $L_3$-edge under  pressure. The positive sign in the energy 
represents 
energy loss. Strong low energy intensities for all high-pressures are mostly attributed to an extrinsic scattering from high-pressure environments such as the Be gasket and the diamond anvils, and those intensity tails largely affect the spectral features below 0.4 eV~\cite{jhkimhp}. The insulating phase (1.8 GPa) spectrum shows two broad features around 0.7~eV and 1.3~eV. At higher pressures, low energy high-pressure environment scattering intensities become stronger, leading to seemingly larger intensities around 0.7 eV, because the gasket and the diamond anvil become closer to the sample at high pressure. The sharp peak around 0.7 eV seen at 3.3 and 5.7~GPa comes from high-pressure environments. On the other hand, the 1.3 eV peak feature is marginally affected by the tail of the extrinsic scattering and free from any sharp high-pressure environment scattering peak. The 1.3 eV peak originates from the orbital excitation in between the occupied $e$ and $a_1$ state and the unoccupied $J_{\rm{eff}}=1/2$ state~\cite{jeong_direct_2017}. The ambient pressure RIXS measurement showed that the 1.3 eV peak intensity is largely modulated with the crystal momentum transfer and the sample angle~\cite{jeong_direct_2017}. The 1.3 eV peak intensity is weak in the sample orientation used for the spectra in Fig.~\ref{RIXS}(a). It is important to note that the 1.3 eV broad feature is, although weak, visible up to the metallic phase (5.7 and 7.1~GPa) and its energy position and width more or less stays the same. For further analysis, see Supplemental Materials \cite{supplementary_materials}.

Figure~\ref{RIXS}(b) presents our main experimental RIXS spectra at the $L_2$-edge under high pressure. In this high-pressure sample, the extrinsic scatterings from high-pressure environments happened to be weaker compared to the case of the $L_3$-edge measurement, and therefore, we resolve orbital excitations above 0.4 eV without high-pressure environment contamination: the low energy extrinsic scattering intensities are similar for all high pressures and no sharp high-pressure environment scattering peak is seen. The 0.7 eV peak at the $L_2$-edge was assigned, in the previous work, to excitations from the occupied $e$ and $a_1$ to the unoccupied $J_{\rm{eff}}=3/2$ states \cite{jeong_direct_2017}. Upon entering the coexistence regime ($P\sim$ 2 GPa; orange), the peak becomes broadened with its intensity reduced. Up to 6 GPa, the peak width and energy are insensitive to the pressure. In the metallic phase ($P=$ 8 and 10.4 GPa; gray and black), the peak width is further broadened. A more itinerant $J_{\rm{eff}}=3/2$ state may contribute to the peak broadening in the high-pressure metallic phase by affecting the local coherent RIXS process. Consistent with the ambient pressure RIXS study, the insulating phase spectrum ($P=$0.9 GPa; red) shows that the 1.3 eV orbital excitation seen at the $L_3$-edge is totally suppressed at the $L_2$-edge due to the destructive quantum interference of the $J_{\rm{eff}}$ state. Importantly, the spectral intensity profile in the 1.3 eV excitation region is insensitive to the applied higher pressure up to 10.4 GPa, confirming that the $J_{\rm eff}$ state persists in the high-pressure metallic phase.

Arguably, this is the first verification of a metallic phase hosting $J_{\rm eff}$=3/2 moments. In the most studied case of a metallic phase derived from a magnetic Mott insulator ({\it e.g.,} cuprates), the magnetic order is quickly destroyed by doping. Recalling that the 
doping of a Mott insulator with $S$=1/2 moments can lead to 
different intriguing phases such as the pseudogap and strange metal phase, or superconductivity, our finding of a metallic $J_{\rm eff}$=3/2 phase deserves further investigations regarding its nature and relation to superconductivity, which are discussed further below.

\begin{figure}[t]
	\includegraphics[width=0.45\textwidth]{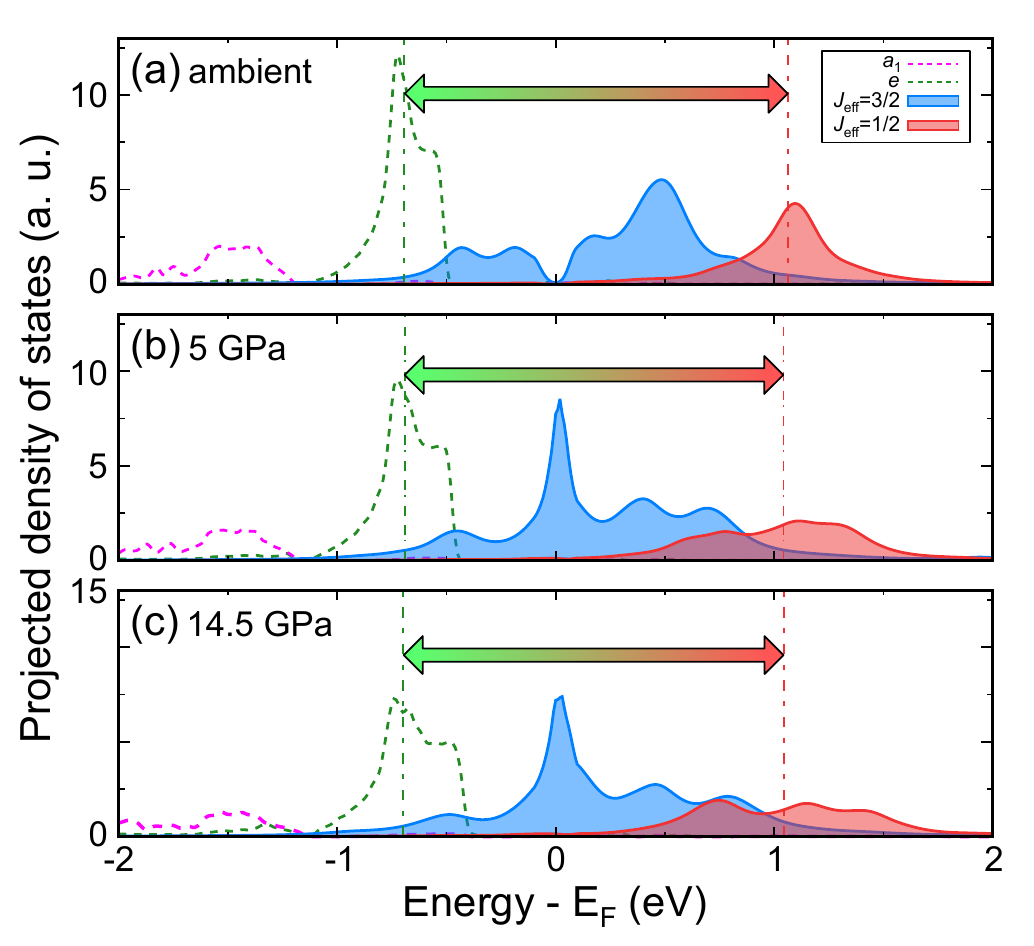}
	\caption{(a-c) Calculated spectral functions at (a) $P=$ 0, (b) 5 and (c) 14.5 GPa. The filled blue and red solid  lines represent the molecular $J_{\rm{eff}}=3/2$ and $J_{\rm{eff}}=1/2$ states, respectively. The dashed magenta and green lines present the $a_1$ and $e$ states, respectively. The arrows connect the center of mass positions of the $J_{\rm{eff}}=1/2$ and $e$ states.
	}
	\label{DOS_pressure}
\end{figure}

{\it Metallic J$_{\rm eff}$=3/2 states: DFT+DMFT calculation --}
In order to further elucidate the characteristics of this novel metallic phase, we performed many-body electronic structure calculations. The DFT+DMFT spectral functions are presented in Fig.~\ref{DOS_pressure}. At ambient pressure (Fig.~\ref{DOS_pressure}(a)), the Mott gap is clearly observed and the upper/lower Hubbard bands are of $J_{\rm{eff}}=3/2$ character. The gap size of 0.4--0.6 eV is in good agreement with optical conductivity data \cite{ta_phuoc_optical_2013}.
At $P\geq$ 5 GPa (Fig.~\ref{DOS_pressure}(b) and (c)), the gap is closed and the system becomes metallic, with a  characteristic quasiparticle peak forming at $E_F$. It should be noted that this correlated metallic feature of the spectral function cannot be captured by the static approximation. See Supplemental Materials for more details \cite{supplementary_materials}.

An important observation is that the center of mass position of the higher-lying $J_{\rm{eff}}=1/2$ states does not move but remains basically unchanged,
even though the 
spectral weight of the $J_{\rm{eff}}=1/2$ states is significantly redistributed by varying pressure.
 The arrows in Fig.~\ref{DOS_pressure} connect the center of mass positions of the $e$ (dashed green lines) and $J_{\rm{eff}}=1/2$ (red solid lines) states,
 and their length is almost independent of pressure. 
 This is particularly important because the $J_{\rm{eff}}=1/2 \rightarrow e$ transition is mainly responsible for the $L_3$ peak at +1.3 eV observed in our RIXS measurement (see Fig.~\ref{RIXS}(a)) \cite{jeong_direct_2017}. Therefore, our DMFT calculation (including SOC) strongly supports our interpretation of the RIXS spectra; namely, the 1.3 eV peak should persist even for the reshaped spectral functions in the metallic phase \cite{comment1}. An additional supporting analysis can be found in Ref.~\onlinecite{supplementary_materials}.

Another noticeable feature is that the low energy states (forming the `coherent peak') in the metallic phase are still of $J_{\rm{eff}}=3/2$ character, see Fig.~\ref{DOS_pressure}(b) and (c). This may have important implications for superconductivity. Recalling the cuprate phase diagram, for example, the Mott insulating state with antiferromagnetic spin order is destroyed by doping and followed by a pseudogap phase before superconductivity appears at low temperature. At higher temperatures, the pseudogap state is followed by the so-called strange metal, whose characteristics are clearly distinct from a Fermi liquid. 
In studies of the two-dimensional Hubbard model, it has recently been shown that these non-Fermi liquid phases host long-lived composite spin-1 moments \cite{werner_spin-freezing_2016,Werner2019}.
In this regard, 
identifying the $J_{\rm{eff}}=3/2$ nature of the metallic phase of GaTa$_4$Se$_8$ may be relevant for understanding the superconductivity observed at higher pressures.

\begin{figure}[t]
	\includegraphics[width=0.45\textwidth]{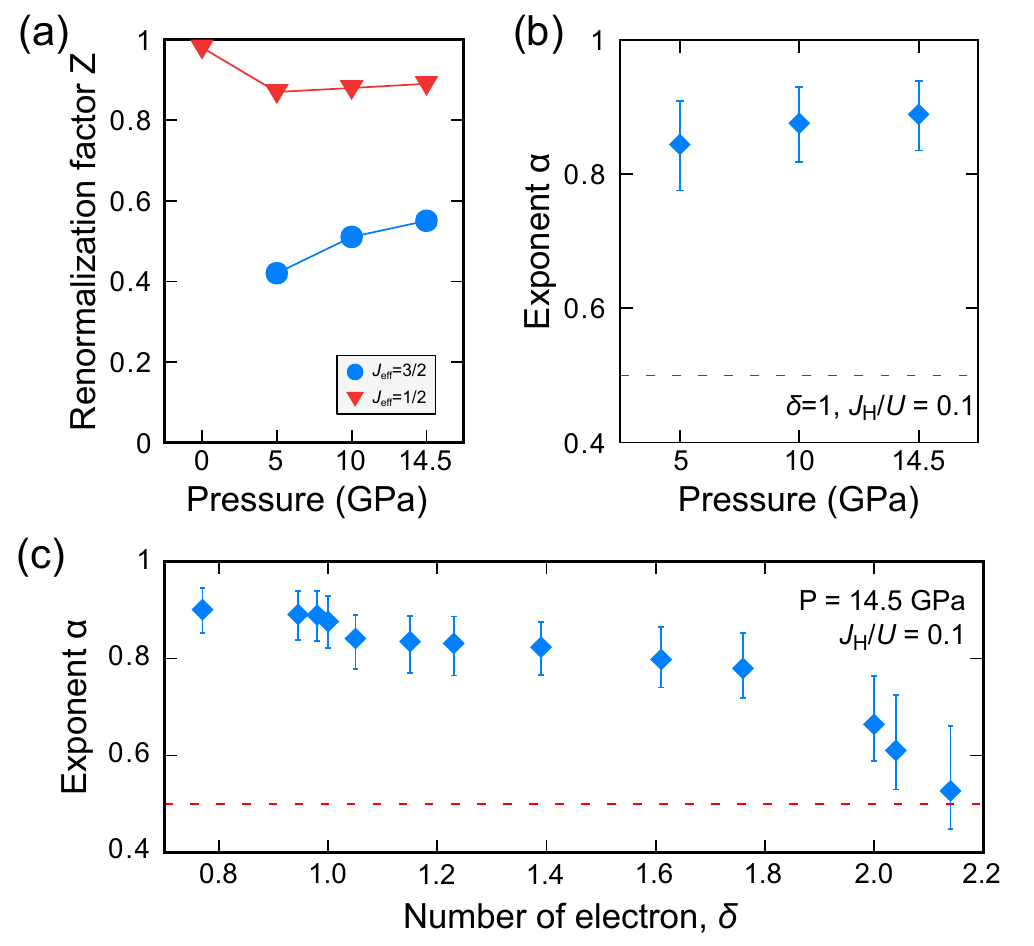}
	\caption{(a) Calculated renormalization factor $Z$ as a function of pressure. Blue circles (red triangles) show the $Z$ of the $J_{\rm{eff}}=3/2$ ($J_{\rm{eff}}=1/2$) bands. (b, c) The calculated exponent $\alpha$ as a function of (b) pressure and (c) electron number. The horizontal red dashed lines show $\alpha=0.5$, namely, the exponent value typically associated with a spin-freezing crossover \cite{werner_spin_2008}. The density of electrons $\delta=1$ corresponds to pristine \GTS. For the exponent fitting, $J_{\rm{H}}/U=0.1$ is used with $U$=0.8 eV.}
	\label{renomalization_factor}
\end{figure}

To gain further insights into the character of this novel metallic phase, we perform a self-energy analysis. The renormalization factor $Z$, defined as $\lim_{\omega\rightarrow 0}[1-\frac{\partial}{\partial \omega}\rm{Re}\Sigma(\omega)]^{-1}$, shows that this pressure-induced phase exhibits sizable electronic correlations, which is reminiscent of the pseudogap or strange metal region of cuprates. Figure~\ref{renomalization_factor}(a) shows that $Z_{J_{\rm eff}=3/2}$ (blue circles) is well below 1.0 while it gradually increases as a function of pressure. This is in contrast to the result for the  $J_{\rm{eff}}=1/2$ bands (red triangles) whose $Z$ value remains close to unity in a wide pressure range.

For cuprates, the relation between the pseudogap phase and superconductivity has long been a central topic of research \cite{norman_pseudogap_2005,lee_doping_2006,keimer_physics_2017,scalapino_common_2012,Gull2013,Werner2019}. Also, recent theoretical studies on half Heusler alloys suggest possible superconductivity arising from a $J=3/2$ band structure \cite{brydon_pairing_2016,timm_inflated_2017,kim_beyond_2018,venderbos_pairing_2018,sim_topological_2019}. Here it is presumed that the observed superconducitivity at higher pressure is unconventional since it emerges out of a novel correlated metallic phase with $J_{\rm{eff}}=3/2$ in proximity to a Mott insulator. 

{\it Doping and spin-freezing superconductivity --}
Finally, we explore and suggest another intruiging possibility in this material. Recent multi-band DMFT calculations showed that unconventional superconductivity can arise from a so-called spin-freezing crossover \cite{werner_spin_2008,hoshino_superconductivity_2015,kim_$mathbfj$_2017,hoshino_superconductivity_2015,werner_nickelate_2020,werner_spin-freezing_2016} although this mechanism still requires experimental confirmation. In order to check this scenario in the case of GaTa$_4$Se$_8$, we perform a self-energy analysis. Following Ref.~\onlinecite{werner_spin_2008}, $-\mathrm{Im}\Sigma(i\omega_n)$ is fitted in the low-energy region with the function $\Gamma + C\cdot(\omega_n)^\alpha$, where $\Gamma, C,$ and $\alpha$ are constants, and $\omega_n$ 
denotes 
Matsubara frequencies. A Fermi liquid is characterized by $\mathrm{Im}\Sigma(i\omega_n) \sim \omega_n$; namely, $\Gamma\approx0$ and $\alpha\approx 1.0$ \cite{georges_strong_2013,kowalski_state_2019,stadler_hundness_2019}. In the moment freezing regime, on the other hand, the self-energy behavior clearly deviates from this linear dependence \cite{georges_strong_2013,kowalski_state_2019}, with $\alpha<1$.

The calculated $\alpha$ is presented as a function of pressure in Fig.~\ref{renomalization_factor}(b) with $J_{\rm{H}}/U=0.1$ where we considered the range of $0.1 \lesssim \omega_n/D \lesssim 0.3$ ($D$ is the half bandwidth) for the fitting. The error bars reflect the deviations caused by varying the fitting range \cite{supplementary_materials}. In the metallic regime of $P\geq$ 5 GPa, $\alpha$ increases as a function of pressure, which is reasonable in the sense that the system becomes more metallic or closer to a Fermi liquid. Note that $\alpha$ is well above the value $\alpha=1/2$ typically associated with the spin freezing crossover, 
which may indicate that the known pressure-induced superconductivity is not primarily driven by local moment fluctuations.
In fact, in this $\delta=1$ system with one electron per $t_2$ molecular orbital, the role of the Hund's interaction $J_{\rm{H}}$ likely becomes less pronounced \cite{georges_strong_2013,stadler_hundness_2019},
although a $J$-freezing crossover has been reported in model calculations \cite{kim_$mathbfj$_2017} with SOC.

On the other hand, we clearly find that introducing extra charges induces a moment freezing crossover in GaTa$_4$Se$_8$. Figure~\ref{renomalization_factor}(c) shows the calculated exponent $\alpha$ as a function of $\delta$ (the electron number per $t_2$ molecular orbital). While $\Gamma$ remains quite small, $\alpha$ is gradually decreased as $\delta$ increases. In particular, at around $\delta\approx$ 1.8--2.0, a substantial drop is observed, indicative of a spin freezing crossover \cite{werner_spin_2008}. This result is also consistent with the previous model study on a Bethe lattice \cite{kim_$mathbfj$_2017}. 
Thus, an unconventional type of superconductivity, possibly distinct from the pressure-induced superconductivity at zero doping, can be expected to occur under electron doping.

In order to introduce extra electrons into GaTa$_4$X$_8$, the chemical substitution of Ge for Ga can be considered, and has been already reported for (Ga/Ge)V$_4$S$_8$ \cite{janod_negative_2015}. Doping alkali- or alkaline-earth metals is another possible way to achieve a spin-freezing crossover. As a `deficient' spinel structure ({\it i.e.,} AB$_2$X$_4$ spinel with half-deficient A sites), lacunar spinels can likely host additional alkali- or alkaline-earth metals. While a spin freezing crossover has been previously suggested for multi-band transition-metal perovskite oxides \cite{georges_strong_2013},
it is awaiting experimental confirmation. Here we note that this prediction is based on the idealized model density of states (DOS) of the Bethe lattice. While the spin freezing crossover appears in between two extreme limits of spin states, this idealized DOS shape can easily be broken up in a real material. Then the system is driven to more stable ordered phases such as antiferromagnetic, ferromagnetic and/or orbital ordered phases, rather than the less stable superconducting phase. This may be the reason why in many multi-band perovskite oxides no superconductivity has been identified. In this regard, GaTa$_4$X$_8$ can be an interesting playground because its DOS shape is better retained due to its molecular nature. Namely, even under pressure, the lattice degree of freedom is less active and the electronic degeneracy is well maintained. In fact, the main change of the lattice structure as a function of pressure is the reduction of the inter-cluster distance, while the molecular units are largely unchanged \cite{pocha_crystal_2005}. Thus GaTa$_4$Se$_8$ can be an ideal platform to explore this type of unconventional superconductivity.

{\it Summary --}
We demonstrated that the metallic phase of GaTa$_4$Se$_8$ carries $J_{\rm eff}$=3/2 moments and exhibits sizable correlations. Our RIXS spectra clearly show that the characteristic orbital excitation features are well maintained under pressure, which is consistent with the results of our DFT+DMFT calculations. The pressure-induced phase can therefore be regarded as a novel type of
correlated metallic phase. Simultaneously, this conclusion suggests that the superconductivity appearing at higher pressure is likely unconventional. Furthermore, our self-energy analysis indicates that an unconventional type of superconductivity may emerge from a $J$-freezing crossover under electron doping. 
Our results highlight a new material phase that has not been observed before, and provides an exciting new playground for exploring unconventional types of superconducting instabilities.

\acknowledgements{ M.Y.J. and S.H.C. contributed equally to this work. The use of the Advanced Photon Source at the Argonne National Laboratory was supported by the U.S. DOE under Contract No. DE-AC02-06CH11357. M.Y.J., J.-H.S and M.J.H were supported by Basic Science Research Program (2018R1A2B2005204), and Creative Materials Discovery Program through NRF (2018M3D1A1058754) funded by the Ministry of Science and ICT (MSIT) of Korea, and the KAIST Grand Challenge 30 Project (KC30) in 2019 funded by the MSIT of Korea and KAIST.}



\bibliography{dmft_GTSe_200211}

\begin{thebibliography}{62}
\bibitem [{com()}]{comment1}%
  \BibitemOpen
  \href@noop {} {}\bibinfo {note} {This weight transfer might be responsible
  for the reduction of $L_3$ RIXS peak height as a function of pressure (see
  Fig.~\ref{RIXS}(b)) which needs further clarification.}\BibitemShut {Stop}%
\end{thebibliography}

\clearpage

\renewcommand{\figurename}{Supplemental Figure}
\renewcommand{\bibname}{Supplemental References}
\renewcommand{\thesection}{\textbf{Supplemental Section~\arabic{section}}}  
\setcounter{figure}{0}

\section*{Supplmentary Materials}

\section{Computational details}

For the first-principles DFT calculations, we use our `OpenMX' software package based on non-orthogonal pseudoatomic local orbitals \cite{ozaki_variationally_2003}. Within the local density approximation (LDA), a $12\times12\times12$ k-grid and 500 Ry energy cutoff is adopted for calculations with a unit cell containing 13 atoms. Cutoff radii of 8.0, 7.0 and 8.0 a.u. were used for Ga, Ta and Se, respectively. 
As is well known, LDA underestimates the cell volume or equivalently the lattice parameter. In the current case, we found that the LDA-optimized volume is $\sim$3\% smaller than the the experimental one. Recent DFT+DMFT studies show that the inclusion of correlation effects can remedy this problem and well reproduce the experimental lattice values \cite{dmft_lattice1_PhysRevLett.106.106405,dmft_lattice2_PhysRevLett.115.256402,dmft_lattice3_PhysRevLett.115.106402,dmft_lattice4_PhysRevB.96.035137}. In the current study, we used the
experimental lattice structures at four different pressure values (0, 5, 10, and 14.5 GPa) \cite{pocha_crystal_2005}.

In the DMFT calculations, we consider a Hamiltonian of the form
\begin{equation}
H=H_{\rm{0}}+\sum_i (H_{\rm{I},i}+H_{\rm{SOC},i}),
\end{equation}
where the non-interacting $H_{\rm 0}$ for molecular $t_2$ orbitals is constructed from the DFT-LDA band structure via the maximally localized Wannier function technique \cite{marzari_maximally_1997,souza_maximally_2001} with a k-grid of $8\times8\times8$. 
$H_{\rm{I},i}$ and $H_{\rm{SOC},i}$ represent the interaction and spin-orbit coupling terms at lattice site $i$. DMFT maps the lattice system to a quantum impurity model with the same $H_{\rm{I}}$ and $H_{\rm{SOC}}$ and a self-consistently determined bath of noninteracting electrons. As impurity solver, we employ the exact diagonalization (ED) method, which requires a discretization of the bath. For $H_{\rm{I}}$, we use a
Kanamori-type parameterization for the three-fold degenerate molecular $t_2$ orbitals: 
\begin{align}
H_{\rm{I}}= & U\sum_{\alpha}{n_{\alpha\uparrow}n_{\alpha\downarrow}}+U^{\prime}\sum_{\alpha\neq\alpha^{\prime}}{n_{\alpha\uparrow}n_{\alpha^{\prime}\downarrow}} \nonumber\\
&+  (U^{\prime}-J_{\rm{H}})\sum_{\alpha<\alpha^{\prime},\sigma}{n_{\alpha\sigma}n_{\alpha^{\prime} \sigma}} - J_{\rm{H}}\sum_{\alpha\neq\alpha^{\prime}}d^{\dagger}_{\alpha\uparrow}d_{\alpha\downarrow}d^{\dagger}_{\alpha^{\prime}\uparrow}d_{\alpha^{\prime}\downarrow}\nonumber\\
& + J_{\rm{H}}\sum_{\alpha\neq\alpha^{\prime}}d^{\dagger}_{\alpha\uparrow}d^{\dagger}_{\alpha\downarrow}d_{\alpha^{\prime}\uparrow}d_{\alpha^{\prime}\downarrow},
\label{interaction}
\end{align}
where $\alpha, \alpha^{\prime} = 1,2,3$ are orbital indices and $\sigma=\uparrow,\downarrow$ denotes spin. 
$U$ is the intra-orbital interaction, $U'$ the inter-orbital opposite-spin interaction, and $J_{\rm H}$ the Hund coupling.
Considering the cubic symmetry, $U^{\prime}$ is defined as $U-2J_{\rm{H}}$.
The SOC is essential for understanding the electronic properties and novel magnetism in GaTa$_4$Se$_8$ \cite{kim_spin-orbital_2014,jeong_direct_2017}. In order to incorporate its effect together with the electron correlations, we consider the term 
\begin{equation}
H_{\rm{SOC}}=\lambda \vec{l}_{\rm{eff}}\cdot\vec{s}
\end{equation}
in our DMFT calculation where $\vec{l}_{\rm{eff}}$, $\vec{s}$, and $\lambda$ represents the effective orbital angular momentum, spin angular momentum, and the strength of SOC, respectively.
Because of the small value of $J_{H}/U$ ($<0.1$), as confirmed by our cRPA calculation \cite{aryasetiawan_calculations_2006,sasioglu_effective_2011}, we neglect the Hund coupling ($J_H=0$) unless otherwise mentioned.
The effective impurity Hamiltonian is defined with 6 correlated orbitals and 18 bath orbitals (including spin degree of freedom). The previous systematic studies on the bath size of ED calculations show that three bath orbitals per one correlated orbital is sufficient to describe multi-orbital physics \cite{senechal_bath_PhysRevB.81.235125,go_spatial_PhysRevLett.114.016402,go_adaptively_PhysRevB.96.085139,Ferititta_rigorous_PhysRevB.98.235132,Linden_imaginary_time_PhysRevB.101.041101}. In particular, the low-frequency features, which are important for the insulator-metal transition (IMT), do not strongly depend on the bath size \cite{go_spatial_PhysRevLett.114.016402,Linden_imaginary_time_PhysRevB.101.041101}. 
The DMFT self-consistency loop is terminated when the bath parameters remain unchanged (within a given tolerance) after the bath fitting. Due to the computational cost, our calculations are performed at zero temperature except for the spectral functions in Fig.\~4, which are obtained for $T\approx 22.6$ K. 
{A double-counting potential of 0.5 eV was used for Fig. 4, which compares well with the value obtained by another widely-used double-counting form, the `FLL (fully localized limit)', which yields 0.4 eV.}
{As commonly observed in DMFT calculations, the coexistence region in Fig. 2 refers to the area which can be either metallic or insulating depending on the initial input self-energy. This area can be connected to the first-order phase transition and the hysteresis is observed in the resistivity-temperature curve near 3.5 GPa for \GTS~\cite{camjayi_first-order_2014}.}
{The renormalization factor $Z$ was computed with the self-energy on the real-frequency axis $\Sigma(\omega)$. ED-DMFT can access directly quantities such as the Green's functions and self-energies on the real-frequency axis, without analytical continuation. For the investigation of non-Fermi liquid, the critical exponent $\alpha$ is calculated on the imaginary-frequency axis and the details are described in supplemental section 7.}
\\

\section{Experimental Details}

The sample was grown by the vapor transport method in a sealed quartz tube. A pair of 0.14 carat standard design diamond anvils with 500 $\mu$m culets were fixed at the seats of a Mao-type symmetric pressure cell which has four windows with ~30$^\circ$ opening. A piece of \GTS single crystal of 80$\times$80$\times$20 $\mu$m$^3$ size was placed inside a 250 $\mu$m diameter hole of a Be gasket along with ruby chips as pressure reference. After closing the cell, Ne gas was loaded as a pressure medium. 

The RIXS measurements at room temperature were performed using the RIXS spectrometer at the 27-ID beamline of the Advanced Photon Source where the sample, analyzer and detector are positioned in the Rowland geometry~\cite{yuri13}. The diamond(111) high-heat-load monochromator reflects x-rays from two in-line undulators into a high resolution monochromator. A four-bounce MERIX medium resolution monochromator produced x-ray beams for both the Ta L$_3$ (9.879 keV) and L$_2$ (11.133 keV) edges with 70 meV bandpass. The beam is then focused by a set of Kirkpatrick-Baez mirrors, yielding a typical spot size of 15$\times$40~$\mu$m$^2$ full-width-half-maximum (FWHM) (v$\times$h) at the sample. For the L$_3$ (L$_2$)-edge RIXS, a Si 066 (Si 466) diced spherical analyzer with 4 inches radius and a position-sensitive silicon microstrip detector were used. The overall energy resolution of the RIXS spectrometer at both edges was 100 meV, as determined from the FWHM of the elastic peak. A horizontal scattering geometry was used with the incident photon polarization in the scattering plane. The scattering angle is fixed at $90^\circ$ and the x-ray beam is incident (scattered) on (from) the loaded sample through the Be gasket. The pressure was $in$-$situ$ controlled by a membrane system and measured by an online Ruby system.

\section{Extrinsic scattering from high-pressure environments}
\label{l3rixs}

\begin{figure}[h]
	\includegraphics[width=0.45\textwidth]{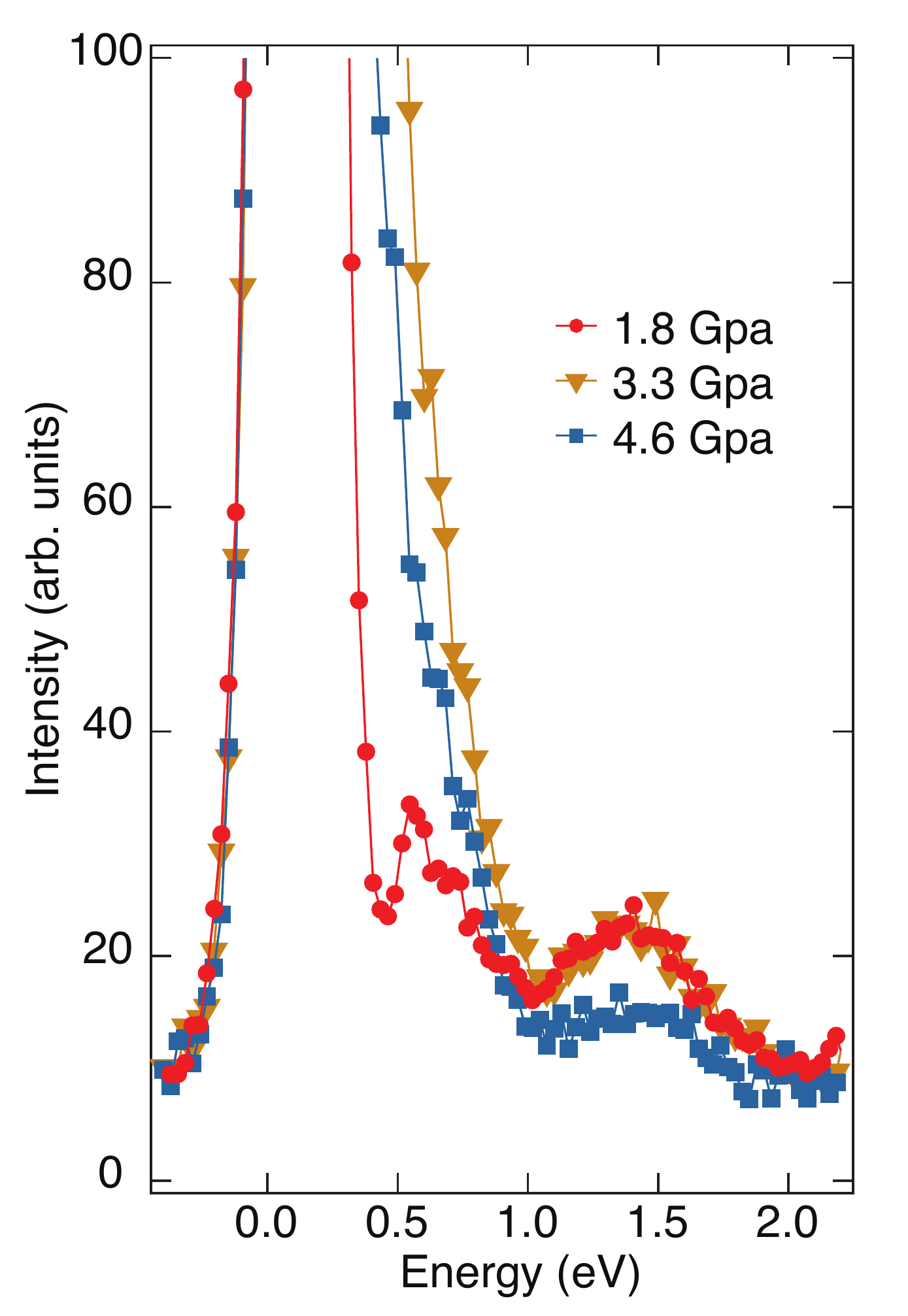}
	\caption{$L_3${-edge} RIXS data collected at a different sample angle from those in Fig. 3(a) plotted on a large scale for a better view of the extrinsic scattering from the high-pressure environment.}
	\label{RIXS_sup}
\end{figure}

Supplemental Figure~\ref{RIXS_sup} shows the pressure-dependent $L_3$ RIXS data on a larger scale than those in Fig.~3(a). These spectra were obtained at a different sample angle from those in Fig.~3(a). The 1.3 eV peak intensity is larger for the sample angle of Supplemental Fig.~\ref{RIXS_sup}, indicating that the 1.3 eV peak intensity is still modulated with the sample angle at high pressure~\cite{jeong_direct_2017}.

In Supplemental Fig.~\ref{RIXS_sup}, the extrinsic elastic scatterings from high-pressure environments such as the Be gasket and the diamond anvils exist as a finite scattering at positive energy as well as a strong quasi-elastic intensity. All measurements in this work were carried out using the diced spherical analyzer with the position-sensitive energy resolving detector~\cite{yuri13}. The diced spherical analyzer scheme lacks a confocal capacity and the detector image of a point source is twice as large as the diced size. So the high-pressure environment scatterings are unavoidable in the low energy region. When the scattering source is located in a different position, the elastic scattering from this source hits a different detector position and is recorded as a finite energy scattering~\cite{jhkimhp}. Because an extrinsic scattering source is nearby the sample and a focused x-ray is used, the high-pressure environment scatterings show mostly as a quasi-elastic intensity and are not observed in the high energy region~\cite{jhkimhp,rossi09}. At 3.3 and 4.6 GPa, the extrinsic scatterings around 0.6 eV become larger compared to the case at 1.8 GPa but the 1.3 eV peak intensity is not changed or decreased, indicating that the 1.3~eV peak feature is free from high-pressure environment scatterings.

In addition, the Si (066) diced analyzer of the Ta $L_3$-edge collects the $\pm$650 meV energy window at the fixed angle. Then the extrinsic elastic scattering is only visible at 1.3 eV when the analyzer center energy is at 650 meV energy loss. Except for this singular energy point, the extrinsic elastic scattering cannot be mapped on the 1.3 eV energy loss position of the position sensitive detector. Therefore, the extrinsic elastic scattering unlikely contributes to the 1.3 eV energy region. In the case of the Si (664) diced analyzer of the Ta L$_2$, it collects the wider energy range of $\pm$1000 meV at a fixed angle and is more susceptible to the extrinsic elastic scattering contribution if it exists. Indeed, if the 1.3 eV peak of $L_3$ originates from extrinsic effect, it should be found in the $L_2$-edge spectra, too. However, as shown in Fig. 3, such an extrinsic elastic scattering contribution is absent in the $L_2$-edge RIXS spectra.

\section{Peak positions in $L_3$-edge RIXS spectra}
While the 1.3 eV peak of $L_3$-edge RIXS spectra is identified in the metallic regime, its peak intensity is  reduced (see Fig. 3(a)) which requires further analysis and clarification of the presence, position and the pressure dependence of this characteristic peak. The results of Gaussian fitting are presented in Supplemental Fig.~\ref{RIXS_fitting_pressure}. While the peak intensity varies, the peak at $\sim$1.3 eV is well identified in the entire pressure range we considered. The possibility of the extrinsic effect or origin of this peak can safely be excluded as discussed in Supplemental Section 3. Importantly, the positions and FWHM values of both peaks at $\sim$0.65 eV and $\sim$1.3 eV are largely unchanged as pressure varies (see Supplemental Fig. \ref{RIXS_peakposition}(b) and (c)). This experimental observation is consistent with the DMFT spectral function shown in Fig.~4. The systematic understanding of the intensity  reduction by pressure is much more challenging not only because there can be several different possible sources to induce it but also because the investigation of this issue requires further experimental and instrumental developments. For example, here we note that the peak intensity can be dependent on the crystal momentum transfer and the sample angle  as already observed at ambient pressure \cite{jeong_direct_2017}. Presumably, the same effect is also present in the pressurized sample. While the systematic and quantitative analysis of the momentum transfer and the angle dependence is strongly requested to address this issue, our current RIXS measurement was mainly designed to verify the absence of the 1.3 eV peak in the $L_2$ at high pressure, and such a momentum transfer was not tracked down at each pressure. We leave further analysis as a furture work.

\begin{figure}[ht]
	\includegraphics[width=0.45\textwidth]{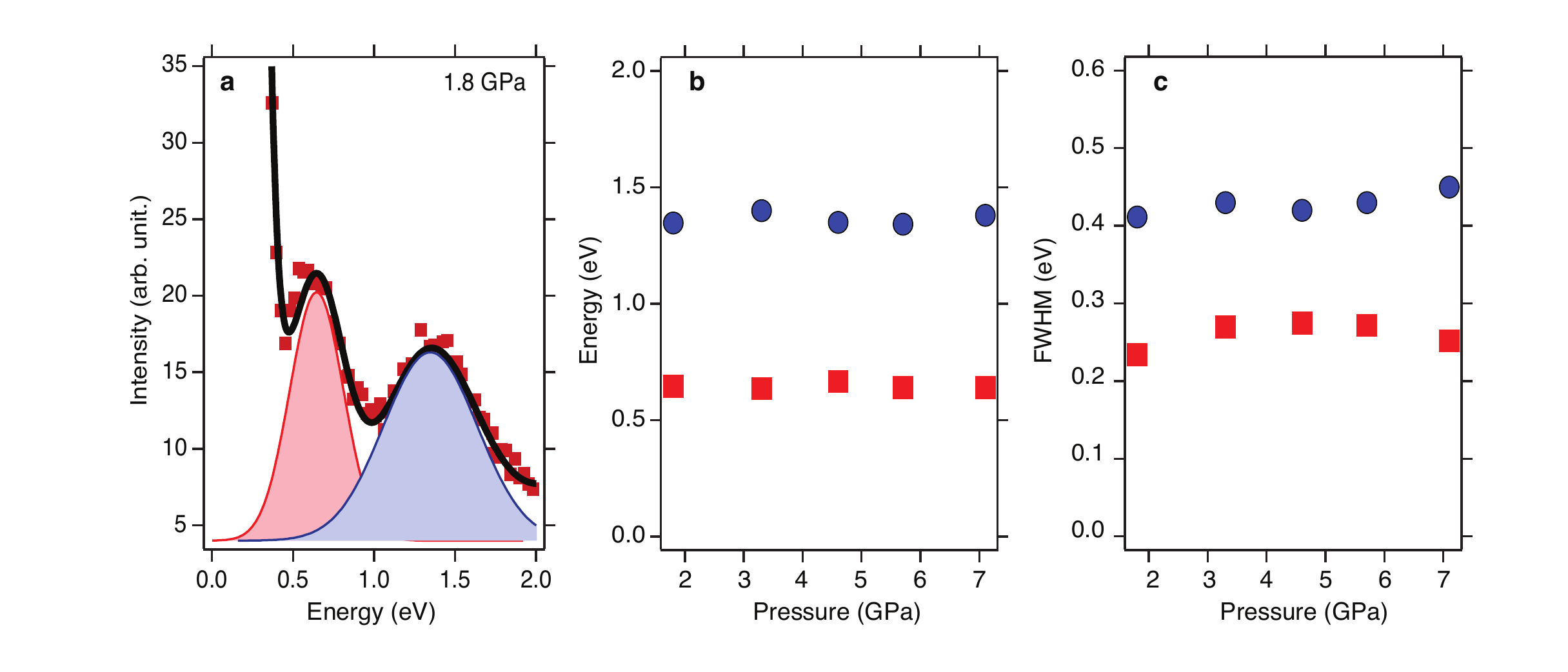}
	\caption{The pressure dependent RIXS spectra at $L_3$-edge. (a) RIXS spectrum at 1.8 GPa and the evolution of (b) the positions of two peaks and (c) FWHM values of both peaks at $\sim$0.65 eV (red) and $\sim$1.3 eV (blue).}
	\label{RIXS_peakposition}
\end{figure}

\begin{figure}[ht]
	\includegraphics[width=0.45\textwidth]{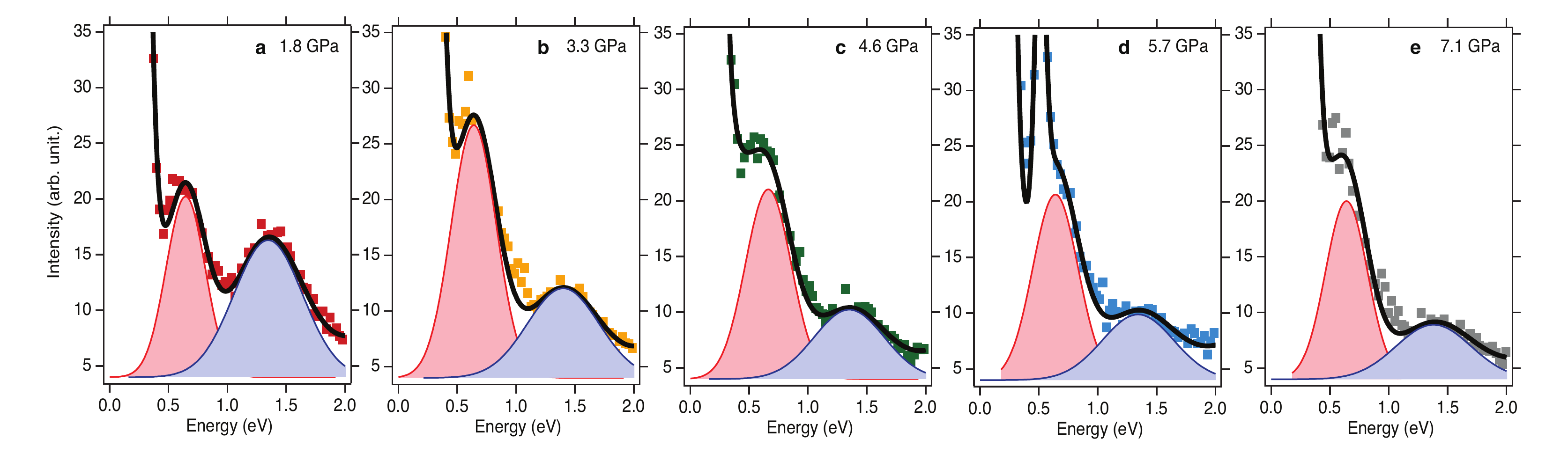}
	\caption{Further fitting results of the pressure dependent RIXS spectra measured at $L_3$-edge.}
	\label{RIXS_fitting_pressure}
\end{figure}

	\section{LDA and LDA$+U$ calculations}
	\label{lda_ldau}
	For the purpose of comparison, we also performed LDA$+U$ calculations which, by construction, favor local moment formation and are limited in the description of correlated metallic phases. It is well established that LDA$+U$ can be regarded as a static Hartree-Fock approximation of LDA+DMFT \cite{kotliar_electronic_2006}. The calculated DOS at ambient pressure ($P=0$) and $P = 14.5$ GPa are presented in Supplemental Fig.~\ref{LDA_LDAU_dos} (a) and (b), respectively. While the $P=0$ result compares reasonably well with the DMFT spectral function shown in Fig. 4(a),
	the high pressure result is significantly different. First of all, the coherent band at $E_F$ is absent in the LDA$+U$ result.
	The DOS shape is quite well maintained in LDA$+U$ and the spectral weight redistribution is much reduced compared to DMFT. Interestingly, the system becomes metallic under pressure in the sense that a small density of states states exists at $E_F$ which is the consequence of an increased bandwidth. Here we used $U=3$ eV for the Ta atom in the LDA$+U$ calculations. It should be noted that this $U$ represents the atomic Ta value, and is therefore significantly greater than that of the molecular $t_2$ orbital \cite{jeong_direct_2017}. We tried to estimate the atomic $U$ value for Ta in this system by using the cRPA method, and found serious convergence problems, which 
	may be due to the 
	molecular nature of the Ta$_4$ cluster. A rough estimation is that the $U$ value for the molecular $t_2$ orbitals should be about 25\% of the atomic Ta value \cite{kim_molecular_2018}. This motivates the use of $\sim 3$ eV in the LDA$+U$ calculations.

	\begin{figure}[h]
		\includegraphics[width=0.45\textwidth]{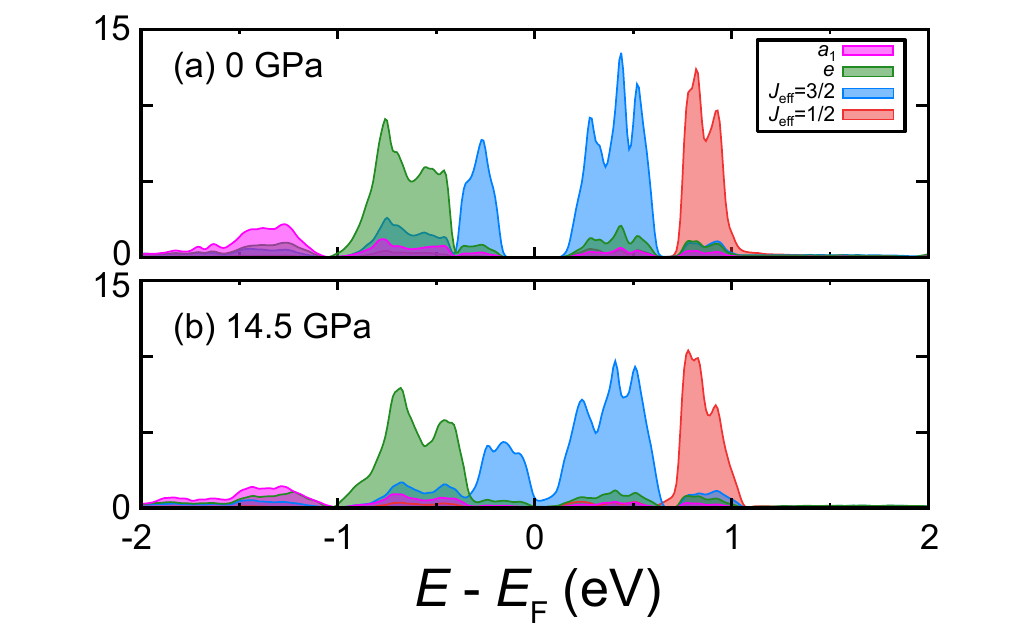}
		\caption{(a, b) Projected DOS calculated by LDA$+U$ under (a) ambient pressure and (b) 14.5 GPa. Violet, green, blue and red lines represent the $a_1$, $e$, $J_{\rm eff}$=3/2 and 1/2 states, respectively.
		}
		\label{LDA_LDAU_dos}
	\end{figure}
	
	\section{Transition probability -- further analysis}
	\label{transit_probability}

	One interesting finding of the current work is that the characteristic RIXS peak only available at $L_3$ (forbidden at $L_2$) persists around +1.2 eV in the metallic regime, just as in the Mott insulating phase at ambient pressure (Fig.~3). This is particularly interesting because our DFT+DMFT spectral functions show substantial spectral weight redistribution across the insulator-metal phase boundary (Fig.~4). In the main text we argue that this can be attributed to the almost pressure-independent position of the center of mass of the $J_{\rm eff}$=1/2 states. Here, we provide an additional analysis on this issue. In Supplemental Fig.~\ref{dostransit}, we plot the approximate transition probability between the $e, a_1$ states and the $J_{\rm eff}=1/2$ states, which is responsible for the $L_3$ RIXS peak. Using the projected DOS, the simple estimate is defined as follows:
	\begin{equation}
	P(\Delta E)=\int d\varepsilon_1 d\varepsilon_2 ~ {D_{o}(\varepsilon_1)D_{u}(\varepsilon_2)} \delta(\varepsilon_2-\varepsilon_1=\Delta E), 
	\end{equation}
	where $D_{o}$ and $D_{u}$ represents the DOS for the corresponding occupied and unoccupied states. While this estimate is not identical with the RIXS spectrum, it provides an informative guide for the peak positions. Supplemental Figure~\ref{dostransit} shows that, across the Mott insulator to metal transition, the center position of the main peak does not change significantly. This result provides additional support for our interpretation of the RIXS spectrum as a function of pressure.

	\begin{figure}[h]
		\includegraphics[width=0.45\textwidth]{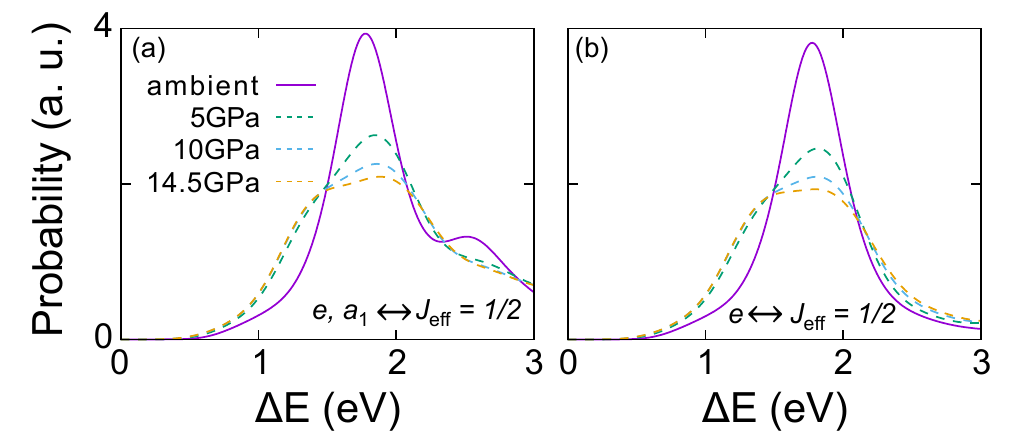}
		\caption{(a, b) Transition probabilities calculated using the projected DOS. For the unoccupied states, two different cases have been considered: (a) both $e$ and $a_1$ and (b) only the $e$ molecular orbital. Purple (solid), green (dashed), cyan (dashed) and orange (dashed) lines represent the result for $P=0$, $5$, $10$ and $14.5$ GPa, respectively. Note that, while the shape of these probability functions changes across the insulator to metal transition ($P \geq 5$ GPa), the main peak position does not change much. This explains why the +1.2 eV peak at the $L_3$ edge persists throughout the measured pressure range. A Gaussian broadening of 0.1 eV has been used in our plot. 
		}
		\label{dostransit}
	\end{figure}

	\section{Details of self the energy fitting}
	\label{fitting}
	
	The spin freezing crossover is a non-zero temperature phenomenon and the self energy recovers its linear dependence near zero frequency \cite{stadler_hundness_2019,georges_strong_2013,georges_strong_2013,kowalski_state_2019,yin_fractional_2012}.
	Since our ED-DMFT calculation is carried out at zero temperature, non-Fermi liquid or spin-freezing behavior is not well described by the lowest Matsubara frequencies. Thus we make use of the fact that high Matsubara frequency self-energies at low temperature show qualitatively the same behavior with that at high temperature, as is known from previous continuous time QMC studies \cite{georges_strong_2013,kowalski_state_2019}.
	A similar approach was adopted in previous low temperature DMFT studies of iron chalcogenides and ruthenates \cite{yin_fractional_2012}. We perform the self-energy fitting with three Matsubara frequencies and a fictitious $\beta=512$. In the pressure range of 0 to 14.5 GPa, $\omega_n/D \sim 0.05-0.28$ with the half-bandwidth $D$ varying from 0.35 to 0.53 eV.
	We conducted three point fittings with $\omega_n, \omega_{n+1}$ and $\omega_{n+2}$, and present the results for $n=5$ as our main data. Other choices of $n=3,4,6,7$ yield the error bars in Fig.~5(b) and (c). For all cases, $\Gamma \approx 0$ ($<10^{-7}$). The results are presented in Supplemental Fig.~\ref{fig_fitting} which clearly shows that the deviations of the dashed line from the solid line at low Matsubara frequencies becomes more pronounced as $\delta$ increases.
	
	\begin{figure}[h]
		\includegraphics[width=0.45\textwidth]{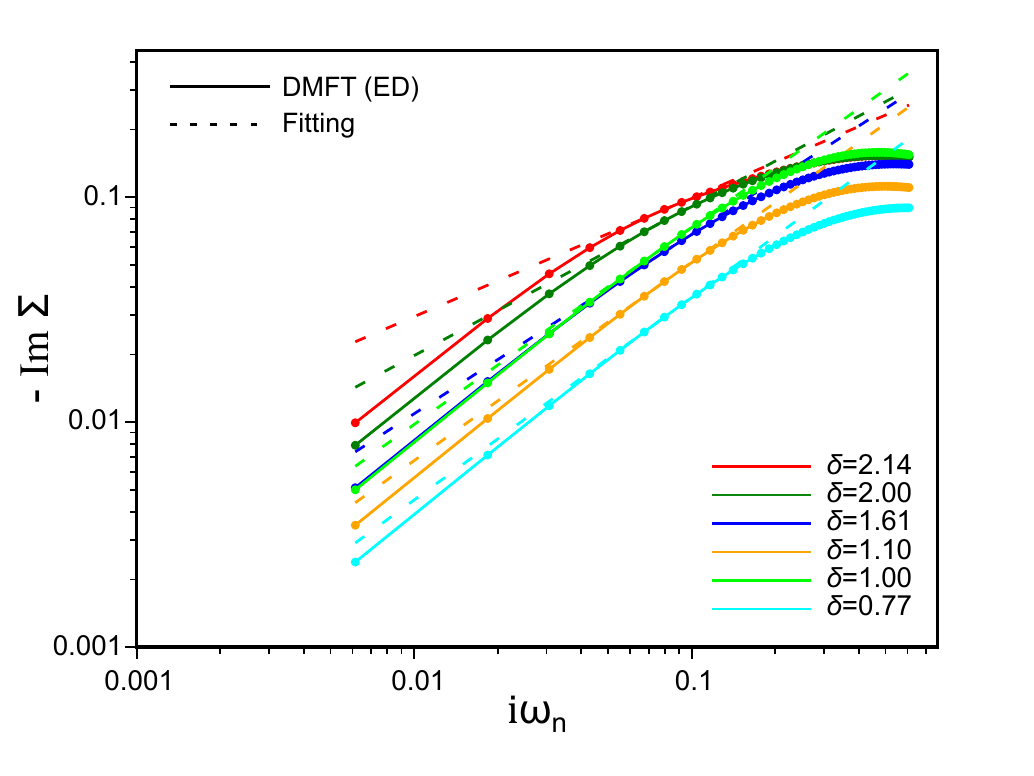}
		\caption{
			Calculated imaginary part of the self energy $-\rm{Im}\Sigma$ on the Matsubara frequency axis. The dashed lines show the fit of the form $\Gamma + C\cdot(\omega_n)^\alpha$ with the fitting parameters $\Gamma, C$, and $\alpha$. Our DFT+DMFT results are presented by circles with solid lines. The cyan, lime, orange, blue, green and red colors represent the electron density $\delta=$0.77, 1.00, 1.10, 1.61, 2.00, and 2.14, respectively.
		}
		\label{fig_fitting}
	\end{figure}


\newpage



\end{document}